\documentclass[aps,twocolumn,floatfix,showpacs,amsmath,amsfonts]{revtex4}

\usepackage{graphicx}
\usepackage{bm}
\usepackage{units}
\usepackage{color}

\begin{document}

\title{Classical dynamics and localization of resonances in the high energy
region of the hydrogen atom in crossed fields}

\author{Frank Schweiner}
\author{J\"org Main}
\author{Holger Cartarius}
\author{G\"unter Wunner}
\affiliation{Institut f\"ur Theoretische Physik 1, Universit\"at Stuttgart,
  70550 Stuttgart, Germany}
\date{\today}

\begin{abstract}
\textcolor{black}{When superimposing the potentials of external fields on the Coulomb 
potential of the hydrogen atom a saddle point appears, which is called the Stark 
saddle point. For energies slightly above the saddle point energy one can 
find classical orbits, which are located in the vicinity of this point. 
We follow those so-called quasi-Penning orbits} to high energies and field strengths observing structural
changes and uncovering their bifurcation behavior. \textcolor{black}{By plotting the stability behavior 
of those orbits against energy and field strength the appearance
of a stability apex is reported. A cusp bifurcation, located in the vicinity of the 
apex, will be investigated in detail. In this cusp bifurcation another orbit of 
similar shape is found, which becomes completely stable in the observed 
region of positive energy, i.e., in a region of parameter space, where the 
Kepler-like orbits located around the nucleus are already unstable}. 
By quantum-mechanically exact calculations we prove the existence of signatures 
in quantum spectra belonging to those orbits. 
Husimi distributions are used to compare quantum-Poincar\'{e}
sections with the extension of the classical torus structure around
\textcolor{black}{the orbits}. Since periodic orbit theory predicts that each classical 
periodic orbit contributes an oscillating term to photoabsorption spectra, we 
finally give an estimation for future experiments, which could 
verify the existence of the stable orbits.
\end{abstract}

\pacs{32.80.-t, 32.60.+i, 05.45.-a}

\maketitle

\section{Introduction}

The hydrogen atom in crossed electric and magnetic fields is a simple
example of a non-integrable physical system, which has up to now been
investigated for almost one hundred years, theoretically \citep{Diss30,Diss31,PRA75_1,PRA75_3,Diss14}
as well as experimentally \citep{PRA75_1,PRA75_10,PRA75_11,PRA75_12,PRA75_3,PRA75_4,PRA75_5,PRA75_6,PRA75_7,PRA75_8,PRA75_9,Diss14}.
Furthermore it has been used for studies on phenomena like Ericson fluctuations
\citep{PRA75_12, PRA75_13} or Arnol'd diffusion \citep{PRA75_10}. 
Even quantum dots \citep{PRA75_15} and excitons \citep{PRA75_14} in condensed matter 
physics can be explored using the findings attained from the hydrogen atom in crossed fields.

One of the major purposes in recent decades was to uncover how chaos,
which can be observed in a classical treatment \citep{MainRob81,MainHar83},
shows itself in quantum spectra, since the Schr\"odinger equation is,
due to its linearity, not capable of producing chaotic behavior \citep{Diss14,MainDel86,MainWin86e,MainWun86b,Diss17,Diss18,Diss5,Diss20}.
Nevertheless new phenomena occur in crossed fields: the so called
quasi-Penning resonances \citep{Diss66}. From a classical point
of view those resonances describe a movement of the electron which
is localized around a saddle point in the potential, the Stark saddle
point. The possible appearance of wave functions localized far away
from the nucleus led to a series of further investigations \citep{Diss66,PRA81_1,PRA81_3,PRA81_5,PRA81_6,MainGay79},
since they can play an important role in the ionization process. Transition
state theory predicts classical orbits localized around the Stark
saddle point and states that all ionizing orbits have to pass the
vicinity of this point \citep{PRA81_7,PRA81_8,PRA81_9,PRA81_10}.
Even though Clark \emph{et al.} \citep{Diss66} already proved their
existence in 1985, it was not until 2009 that first signatures of
those orbits were found in calculated quantum spectra \citep{PRA79}.
The classical stability behavior of the quasi-Penning orbits \textcolor{black}{at high energies and field strengths has
first been investigated by Fl\"othmann in 1994 \citep{Diss}, who uncovered
a stability apex in parameter space, if the stability change of the orbits is illustrated as a function of both parameters.} 

It is the purpose of this paper to perform more precise calculations
on the stability of those orbits and to uncover the bifurcation behavior
as well as the processes taking place around the stability \textcolor{black}{apex}. 
\textcolor{black}{We report the existence of a cusp bifurcation, which appears in the vicinity of the
stability apex and involves another orbit of similar shape.} 
This orbit, becoming completely stable in a specific
area of parameter space, is the starting point for semiclassical and
quantum mechanical calculations. We will demonstrate that signatures
of those orbits at high energies and field strengths can be found
in accurately calculated quantum spectra.

\textcolor{black}{The paper is organized as follows.}
In Sec.~\ref{sec:Classical-calculations} the system is introduced,
and a scaling of parameters as well as a regularization of coordinates
are carried out. A comparison between exceptional points and a classical
cusp bifurcation is drawn. The classical stability of the quasi-Penning
resonances and their bifurcation behavior at high energies and field
strengths are presented in Sec.~\ref{sec:3}. An introduction in the
semiclassical and exact quantum mechanical calculations performed
is given in Sec.~\ref{sec:4}, before according results are discussed
in Sec.~\ref{sec:5}. In Sec.~\ref{sec:6} a short summary is given
and conclusions are drawn.

\section{Classical calculations\label{sec:Classical-calculations}}

\subsection{Hamiltonian, monodromy matrix and regularized coordinates \label{sec:2}}

The \textcolor{black}{Hamiltonian} of a hydrogen atom in a constant electric
field $\boldsymbol{F}=F\hat{e}_{x}$ and a constant magnetic field
$\boldsymbol{B}=B\hat{e}_{z}$ reads in \textcolor{black}{atomic units (Hartree units with lengths, energies, electric and magnetic field strengths given in units of $5.29\times10^{-11}\,\mathrm{m}$, $4.36\times10^{-18}\,\mathrm{J}$, $5.14\times10^{11}\,\mathrm{V/m}$, and $2.35\times10^{5}\,\mathrm{T}$, respectively)}
\begin{equation}
H=\frac{1}{2}p^{2}-\frac{1}{r}+\frac{1}{2}BL_{z}+\frac{1}{8}B^{2}\left(x^{2}+y^{2}\right)+Fx,\label{eq:H}
\end{equation}
with $L_{z}=xp_{y}-yp_{x}$. In the following we shall take advantage
of a scaling property of the \textcolor{black}{classical} Hamiltonian enabling us to deal with
only two independent variables, i.e., the scaled energy and the scaled field
strength. \textcolor{black}{The other variables have to be scaled, as well, so that the scaling transformations read}
\begin{subequations}
\begin{equation}
\tilde{E}=EB^{-{2/3}},\;\tilde{F}=FB^{-{4/3}},
\end{equation}
\begin{equation}
\tilde{\boldsymbol{r}}=\boldsymbol{r}B^{2/3},\;\tilde{\boldsymbol{p}}=\boldsymbol{p}B^{-{1/3}},\;\tilde{t}=tB.
\end{equation}
\label{eq:skal}
\end{subequations}
Without further usage of the tilde sign the Hamiltonian has the
same structure as in Eq.~(\ref{eq:H}) when setting $B=1$. Defining $\boldsymbol{\gamma}=\left(\boldsymbol{r},\,\boldsymbol{p}\right)^{T}$
we want to find periodic solutions of the Hamiltonian equations of
motion
\begin{equation}
\frac{\mathrm{d}}{\mathrm{d}t}\boldsymbol{\gamma}=\boldsymbol{J}\frac{\partial H}{\partial\boldsymbol{\gamma}},\;\mathrm{with}\;\boldsymbol{J}=\left(\begin{array}{cc}
\boldsymbol{0} & \boldsymbol{1}\\
\boldsymbol{-1} & \boldsymbol{0}
\end{array}\right).\label{eq:eqofmotion}
\end{equation}
The Stark saddle point is characterized by the fixed-point condition
$\dot{\boldsymbol{\gamma}}=\boldsymbol{0}$, yielding its position
$\boldsymbol{r}_{\mathrm{SP}}=(-1/\sqrt{F},\,0,\,0)^{T}$ and energy
$E_{\mathrm{SP}}=-2\sqrt{F}$. The stability of orbits is investigated using
the stability matrix $\bar{\boldsymbol{M}}$. $\bar{\boldsymbol{M}}$
describes in a linear approximation the relative behavior of two
trajetories $\boldsymbol{\gamma}^{(1)}$ and $\boldsymbol{\gamma}^{(2)}$
in time \citep{Diss53},
\begin{equation}
\boldsymbol{\gamma}^{(1)}\left(t\right)-\boldsymbol{\gamma}^{(2)}\left(t\right)=\bar{\boldsymbol{M}}\left(t\right)\left(\boldsymbol{\gamma}^{(1)}\left(0\right)-\boldsymbol{\gamma}^{(2)}\left(0\right)\right).
\end{equation}
and can be determined by its equation of motion
\begin{equation}
\frac{\mathrm{d}}{\mathrm{d}t}\bar{\boldsymbol{M}}=\boldsymbol{J}\frac{\partial^{2}H}{\partial\boldsymbol{\gamma}^{2}}\bar{\boldsymbol{M}},\;\mathrm{with}\;\boldsymbol{J}=\left(\begin{array}{cc}
\boldsymbol{0} & \boldsymbol{1}\\
\boldsymbol{-1} & \boldsymbol{0}
\end{array}\right),\;\bar{\boldsymbol{M}}\left(0\right)=\boldsymbol{1}.\label{eq:eqmonodromy}
\end{equation}
The eigenvalues of $\bar{\boldsymbol{M}}\left(t\right)$ therefore
indicate whether a variation of the initial conditions of a periodic
orbit leads to exponential divergence or a trajectory remains in the
vicinity of the periodic orbit for all times. In the case of a Hamiltonian
system the energy conservation leads to two variational directions
not affecting the system's behavior. The corresponding eigenvalues
hence take on the value of $1$. Omitting of those directions one
obtains the \textcolor{black}{so-called} monodromy matrix $\boldsymbol{M}$.
 
To prevent a divergence of the momentum $\boldsymbol{p}$ near the
nucleus the Kustaanheimo-Stiefel regularization of coordinates \citep{Diss47_1,Diss47_2}
is used,
\begin{subequations}
\begin{equation}
\boldsymbol{r}=\boldsymbol{L}\left(\boldsymbol{U}\right)\boldsymbol{U}=\frac{1}{2}\left(\begin{array}{cccc}
U_{3} & -U_{4} & U_{1} & -U_{2}\\
U_{4} & U_{3} & U_{2} & U_{1}\\
U_{1} & U_{2} & -U_{3} & -U_{4}\\
U_{2} & -U_{1} & -U_{4} & U_{3}
\end{array}\right)\boldsymbol{U},\label{eq:reg_r_U}
\end{equation}
\begin{equation}
\boldsymbol{p}=\frac{1}{r}\boldsymbol{L}\left(\boldsymbol{U}\right)\boldsymbol{P},\label{eq:reg_p_P}
\end{equation}
\textcolor{black}{together with a transformation of time,}
\begin{equation}
\mathrm{d}t=2r\,\mathrm{d}\tau.\label{eq:reg_t_tau}
\end{equation}
\label{eq:reg_}
\end{subequations}
The equations of motion (\ref{eq:eqofmotion}) and (\ref{eq:eqmonodromy})
remain the same except for the replacements $\boldsymbol{\gamma}\rightarrow\boldsymbol{\Gamma}=\left(\boldsymbol{U},\,\boldsymbol{P}\right)^{T}$
and $t\rightarrow\tau$. The Hamiltonian then reads
\begin{eqnarray}
H & = & \phantom{+}\frac{1}{2}P^{2}-\left[E-F\left(U_{1}U_{3}-U_{2}U_{4}\right)\right]U^{2}\nonumber \\
 &  & +\frac{1}{2}\left[\left(U_{1}P_{2}-U_{2}P_{1}\right)\left(U_{3}^{2}+U_{4}^{2}\right)\right.\nonumber \\
 &  & \phantom{+\frac{1}{2}}\left.\quad+\left(U_{3}P_{4}-U_{4}P_{3}\right)\left(U_{1}^{2}+U_{2}^{2}\right)\right]\nonumber \\
 &  & +\frac{1}{8}U^{2}\left(U_{1}^{2}+U_{2}^{2}\right)\left(U_{3}^{2}+U_{4}^{2}\right) = 2,\label{eq:reg_H}
\end{eqnarray}
with $U^{2}=\sum_{i=1}^{4}U_{i}^{2}$. By integrating two further
equations along with the Hamiltonian equations of motion one obtains
the periods and actions of the orbits:
\begin{eqnarray}
\frac{\mathrm{d}}{\mathrm{d}\tau}t & = & U^{2},\\
\frac{\mathrm{d}}{\mathrm{d}\tau}S & = & P^{2}+\frac{1}{2}\left[\left(U_{1}P_{2}-U_{2}P_{1}\right)\left(U_{3}^{2}+U_{4}^{2}\right)\right.\nonumber \\
 &  & \qquad\qquad\left.+\left(U_{3}P_{4}-U_{4}P_{3}\right)\left(U_{1}^{2}+U_{2}^{2}\right)\right].\label{eq:reg_S}
\end{eqnarray}
The transformation (\ref{eq:reg_}) is not
bijective. For this reason the inversion,
\begin{equation}
\boldsymbol{U}=\left(\begin{array}{c}
\sqrt{r+z}\,\vphantom{\int_{2}^{3}}\cos\left(\frac{\varphi+\alpha}{2}\right)\\
\sqrt{r+z}\,\vphantom{\int_{2}^{3}}\sin\left(\frac{\varphi+\alpha}{2}\right)\\
\sqrt{r-z}\,\vphantom{\int_{2}^{3}}\cos\left(\frac{\varphi-\alpha}{2}\right)\\
\sqrt{r-z}\,\vphantom{\int_{2}^{3}}\sin\left(\frac{\varphi-\alpha}{2}\right)
\end{array}\right),\label{eq:U_r}
\end{equation}
contains an additional parameter $\alpha$, which, without loss of
generality, we will set to the constant value $\alpha=0$.

\subsection{Cusp bifurcation and exceptional points\label{sub:Cusp-bifurcation-and}}

\begin{figure}[t]
\includegraphics[width=0.49\columnwidth]{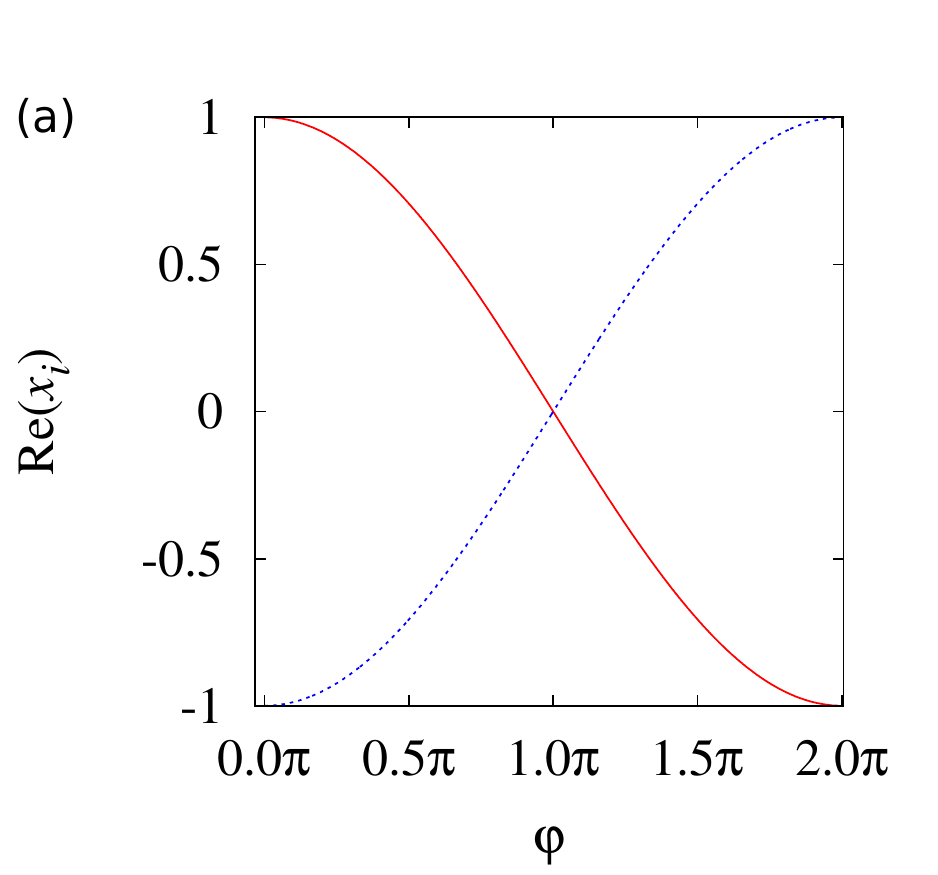}
\includegraphics[width=0.49\columnwidth]{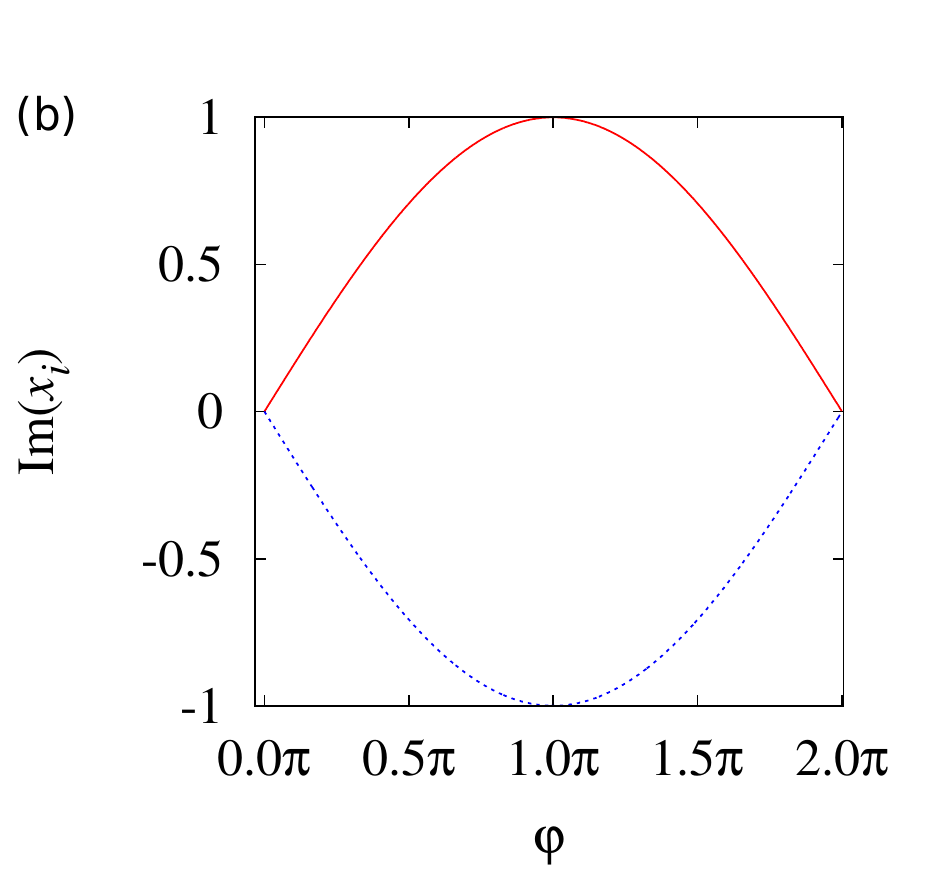}

\includegraphics[width=0.49\columnwidth]{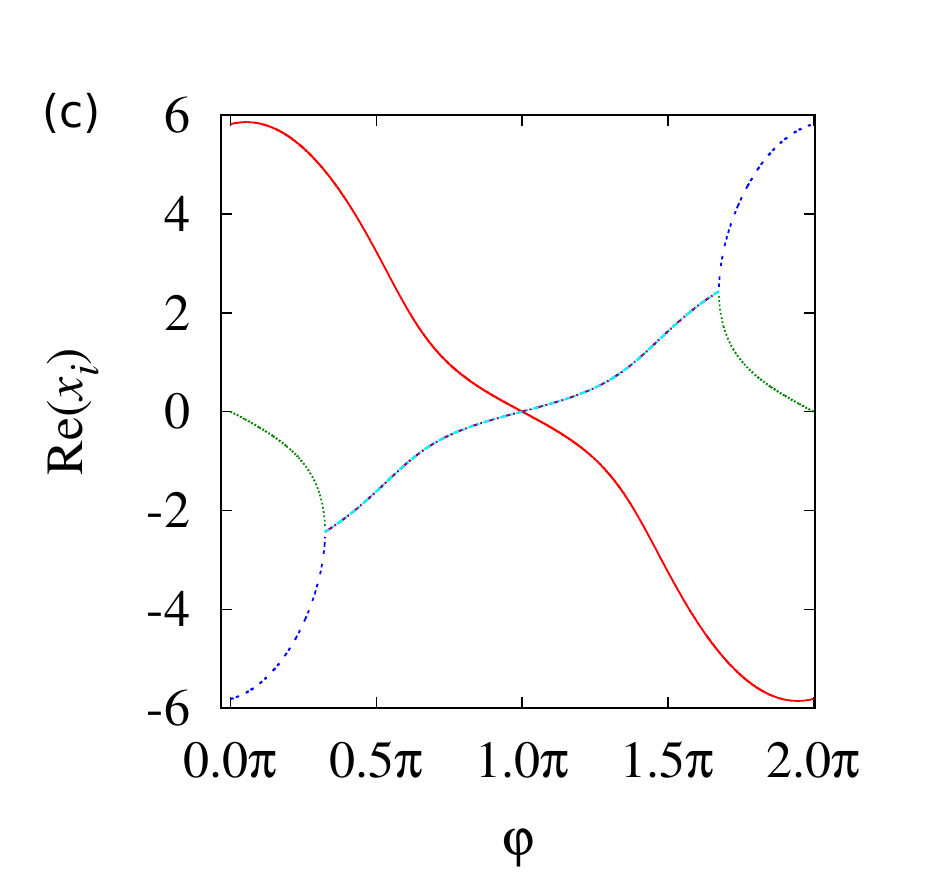}
\includegraphics[width=0.49\columnwidth]{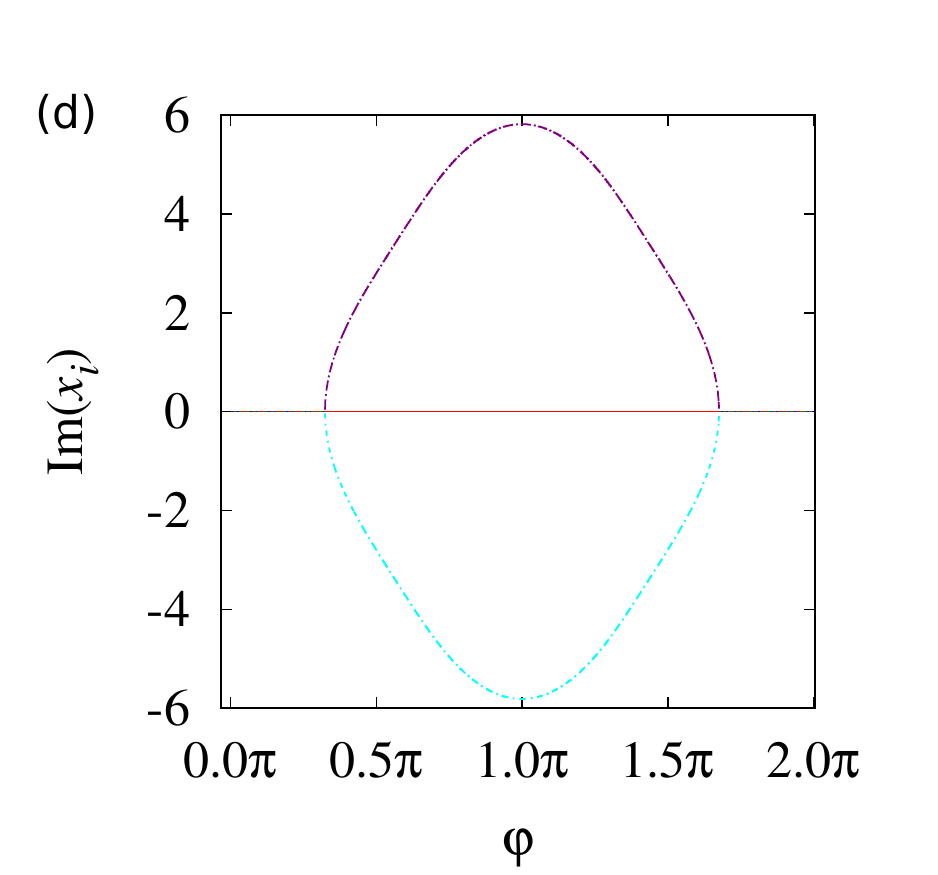}

\caption{(Color online) By encircling an EP$2$ an interchange of two resonances can be observed.
The figures~(a) and (b) show the real and imaginary part of the two fixed
points of Eq.~(\ref{eq:tangent}), depending on the angle $\varphi$
in $\mu=e^{i\varphi}$. 
(c), (d) When encircling
the cusp point along $\left(a,\, b\right)=\left(-5\cos\varphi,\,-5\sin\varphi\right),\,\varphi\in\left[0,\,2\pi\right)$, 
one of the fixed points of Eq.~(\ref{eq:cuspnormal}) can be transformed
into another one (marked by \textcolor{black}{the} red \textcolor{black}{solid} line \textcolor{black}{starting in the upper left corner of figure~(c)}), 
without passing through a bifurcation.
A second encircling passes both tangent bifurcations. Since the fixed
points coincide in those bifurcations, it is not possible to determine
the permutation behavior. An EP$2$- or an EP$3$-behavior can be
observed accordingly. \textcolor{black}{The angle $\varphi$ and the position $x$ are given in dimensionless units (see Eqs.~(\ref{eq:cuspnormal}) and~(\ref{eq:tangent})).}\label{fig:cusp}}
\end{figure}

A cusp bifurcation, appearing in systems with at least two parameters
$a$ and $b$, is described by the normal form \citep{Poston}
\begin{equation}
\dot{x}=\frac{4}{27}x^{3}+ax+b.\label{eq:cuspnormal}
\end{equation}
We choose the factor in front of $x^{3}$ as
$\frac{4}{27}$ instead of $1$ to simplify the results. When calculating
the fixed points of Eq.~(\ref{eq:cuspnormal}) one finds three real
solutions for $a<0$ and $\left|b\right|<\left(-a\right)^{3/2}$.
The two boundary lines of this area are tangent bifurcation lines,
along which two fixed points coincide. In the remaining parameter
space only one real solution can be found. A specific attribute of
the cusp bifurcation shows a similarity to exceptional points: Following
one fixed point around the cusp point, which is located at $\left(a,\, b\right)=\left(0,\,0\right)$,
along a circle 
\begin{equation}
\left(a,\, b\right)=\left(-r\cos\varphi,\,-r\sin\varphi\right),\:\mathrm{with}\:\varphi\in\left[0,\,2\pi\right)\label{eq:path}
\end{equation}
it can be transformed into another one.
This phenomenon appears in the case of exceptional points, too. An exceptional
point, first described by Kato \citep{JPhysA45_3}, marks the coalescence
of at least two resonances (or more precisely: eigenvalues and corresponding
eigenvectors) of a complex Hamiltonian in an at least two-dimensional
parameter space \citep{Robin_9,Robin_10, JPA45, Robin}. Encircling the exceptional
point in parameter space the resonances permute. In the case of an
EP$2$ two resonances interchange and the initial situation can be
restored after two cycles (Fig.~\ref{fig:cusp}a and~\ref{fig:cusp}b). This can be described
by the normal form of a tangent bifurcation
\begin{equation}
\dot{x}=x^{2}-\mu,\label{eq:tangent}
\end{equation}
when choosing the parameter $\mu$ to be complex. The two complex
fixed points can then be interchanged by encircling the exceptional
point $\mu=0$ along the unit circle in the complex plane $\left(\mu=e^{i\varphi},\,\varphi\in\left[0,\,2\pi\right)\right)$.
Similarly, the behavior of an EP$n$ can be described by the normal
form
\begin{equation}
\dot{x}=x^{n}-\mu.
\end{equation}
When encircling the exceptional point the $n$ fixed points
permute and the initial situation is restored after $n$ cycles.

\begin{figure}[t]
\includegraphics[width=0.8\columnwidth]{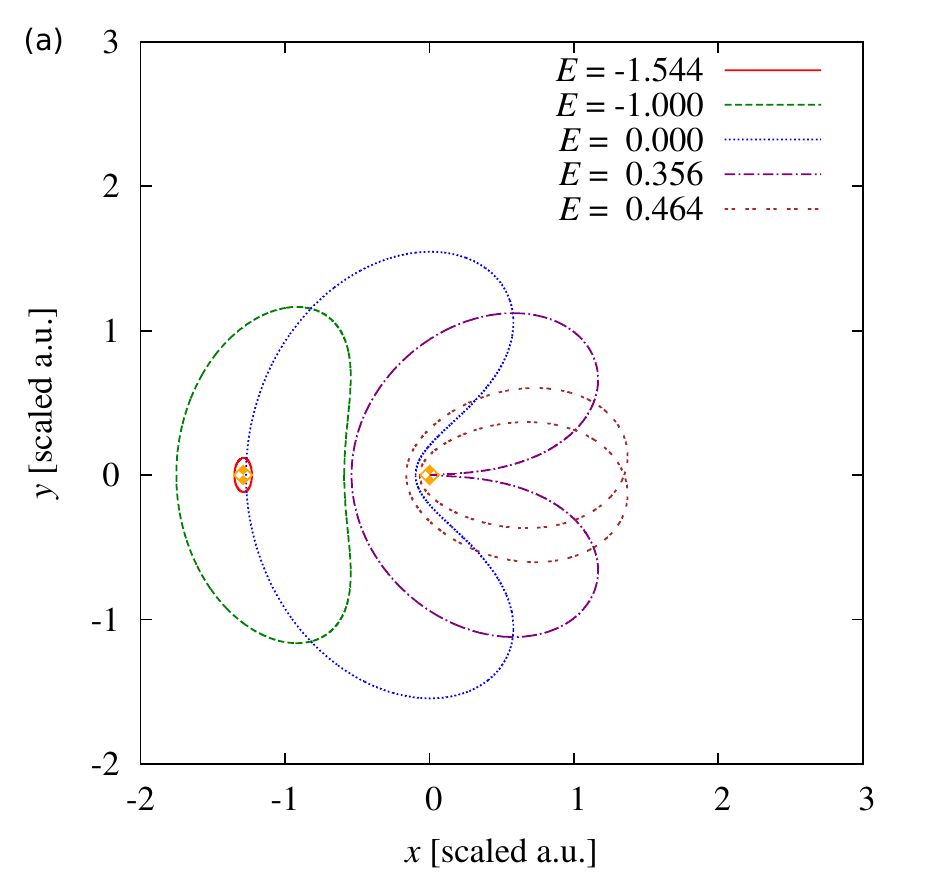}

\includegraphics[width=0.8\columnwidth]{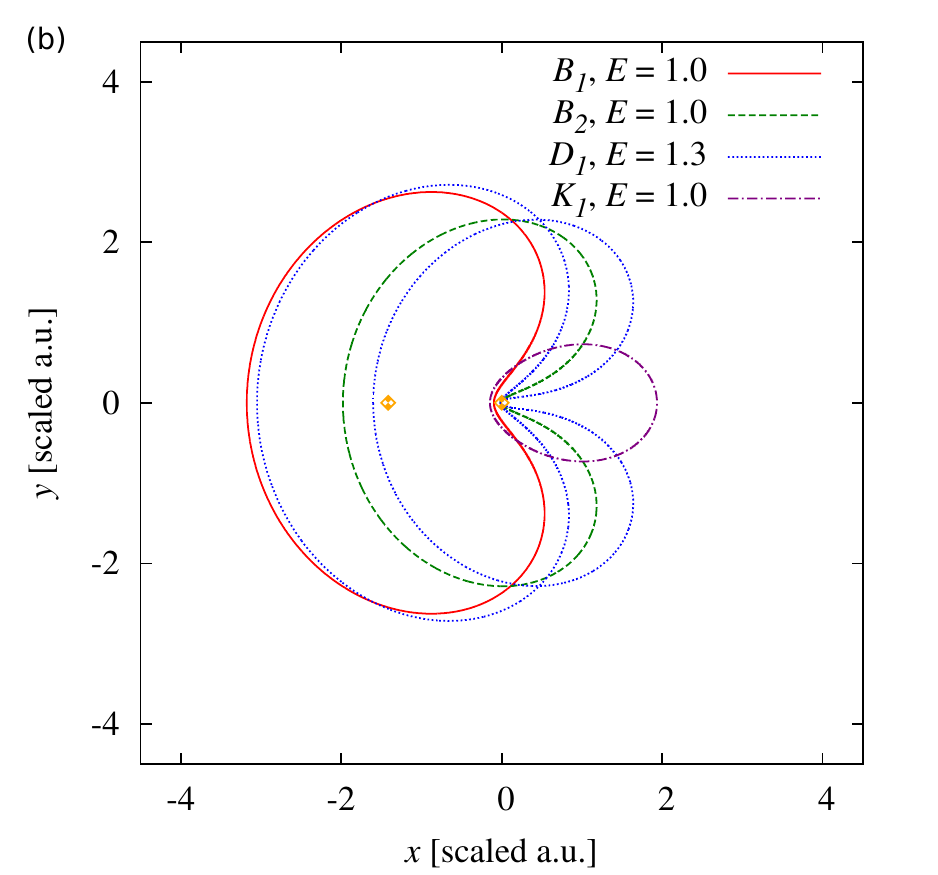}

\caption{(Color online) (a) Shape of the orbit $B_{1}$, evolving
from the elliptically shaped quasi-Penning resonances, at different energies and a fixed
value of $F=0.6$. The orbit is localized
in the $z=0$-plane. The position of the Stark saddle point on the
left hand side and the nucleus at the origin are marked by diamonds.
At $E\approx0.356$ the right point of intersection
with the $x$-axis coincides with the nucleus. 
(b) Shape of the orbits $B_1$, $B_2$, $D_1$ 
and $K_1$ at $F=0.5$. \textcolor{black}{The positions $x$ and $y$ are given in scaled atomic units (see Eq.~(\ref{eq:skal})).}\label{fig:orbits}}
\end{figure}

\textcolor{black}{Depending on the path for the encircling the
cusp bifurcation can show the behavior of an EP$2$ \emph{or} an
EP$3$}: By choosing $b=0$ and $a=e^{i\varphi}$ one obtains an EP$2$-behavior.
On the other hand an EP$3$-behavior is obtained by setting
$a=0$ and $b=e^{i\varphi}$. This phenomenon was already
observed in quantum mechanical calculations for dipolar Bose-Einstein
condensates \citep{JPA45, Robin}. The path of Eq.~(\ref{eq:path}) with
real parameters $a$ and $b$ gives rise to an ambiguity between both
behaviors: If one traces two fixed points beyond a tangent bifurcation,
two complex fixed points will appear. Since those two points coincide
in the tangent bifurcation, it is not possible to relate one of the
complex points to one of the real points, respectively. The permutation
behavior cannot be determined, which is shown in Fig.~\ref{fig:cusp}c and~\ref{fig:cusp}d. 
This ambiguity exists only for real parameter values. 
If a small complex part is added to $a$ or $b$, an unambigious 
determination of the permutation behavior is possible.

\section{Results of classical calculations and discussion \label{sec:3}}

Figure~\ref{fig:orbits}a shows the shape of the quasi-Penning
resonances when following them up to higher energies at a fixed field
strength. Since the name quasi-Penning was applied to almost elliptical
orbits localized in the vicinity of the Stark saddle point and referred
to the structural similarity between the equations of motion linearized
around the saddle point and the stability conditions in a Penning trap
we will now refer to them more generally as $B_{1}$. The differences
to the case of energies slightly above $E_{\mathrm{SP}}$ are obvious:
When increasing the energy the orbit becomes more and more heart-shaped. 
Figure~\ref{fig:orbits}b shows in advance the shortest orbits in the $z=0$-plane,
which are now investigated in detail.

\begin{figure}[t]
\includegraphics[width=0.9\columnwidth]{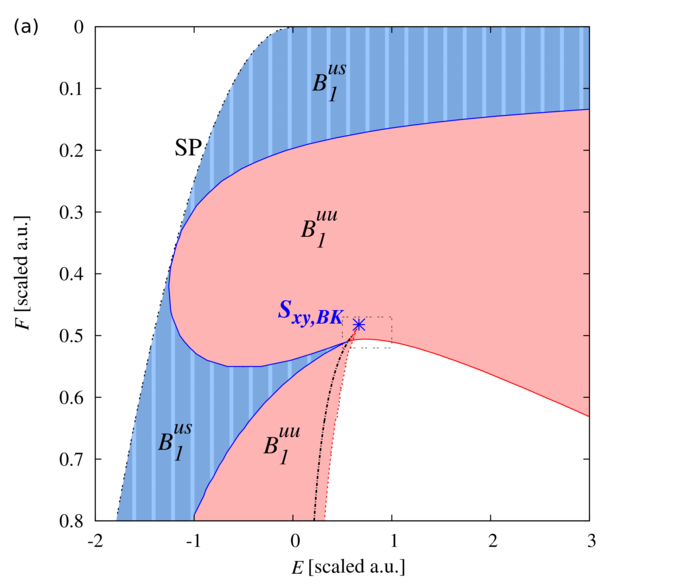}
\includegraphics[width=0.9\columnwidth]{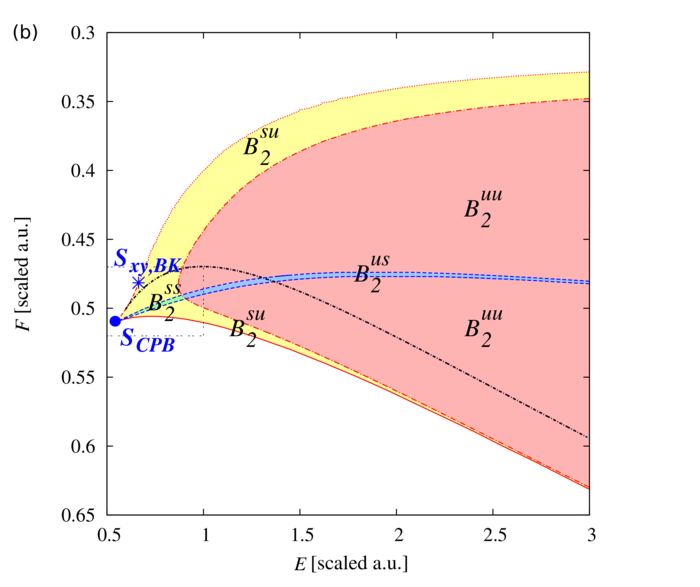}
\caption{(Color online) Stability of $B_{1}$ (a) and $B_{2}$ (b) in dependence on the scaled
energy and the scaled field strength. Along the lines marking the
borders of different stability regions bifurcations occur. The
stability within those regions is displayed by two upper indices.
The first one refers to the stability (s) or instability (u) \textcolor{black}{of the orbit in the direction}
perpendicular to the magnetic field, the second one to the stability \textcolor{black}{in the direction}
parallel to the magnetic field. According to the different
types of stability the regions are additionally dyed by colors\textcolor{black}{, shadings} and grayscales. 
\textcolor{black}{One can clearly observe the stability apex for $B_1$ ending at $F\approx0.5,\, E\approx0.5$, i.e., a point in the $E$-$F$ space at which two border lines of the stability regions $B_1^{us}$-$B_1^{uu}$ meet, which was first uncovered by Fl\"othmann \citep{Diss}.}
The dashed line on the left hand side in (a) marks the energy of the saddle
point (SP) as a function of the field strength. Specific points in parameter
space are displayed by $\boldsymbol{S}_{i}=\left(F_{i},\, E_{i}\right)$. 
\textcolor{black}{The field strength $F$ and the energy $E$ are given in scaled atomic units (see Eq.~(\ref{eq:skal})).}
\label{fig:B1B2}}
\end{figure}

\begin{figure}[t]
\includegraphics[width=0.9\columnwidth]{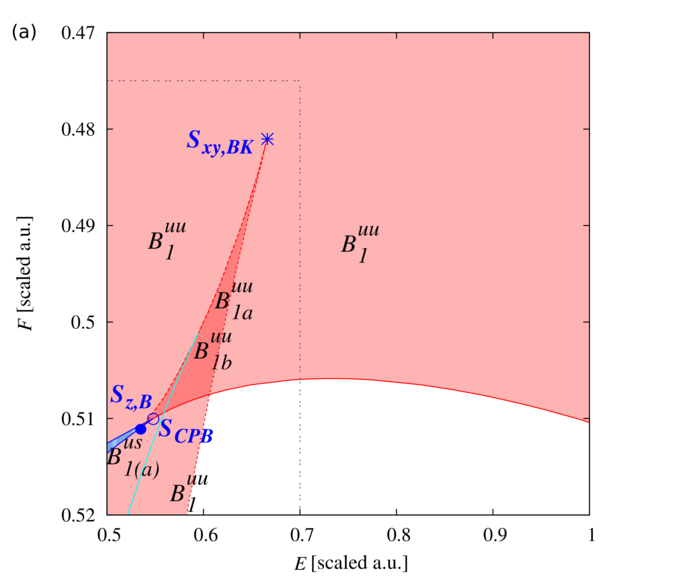}
\includegraphics[width=0.9\columnwidth]{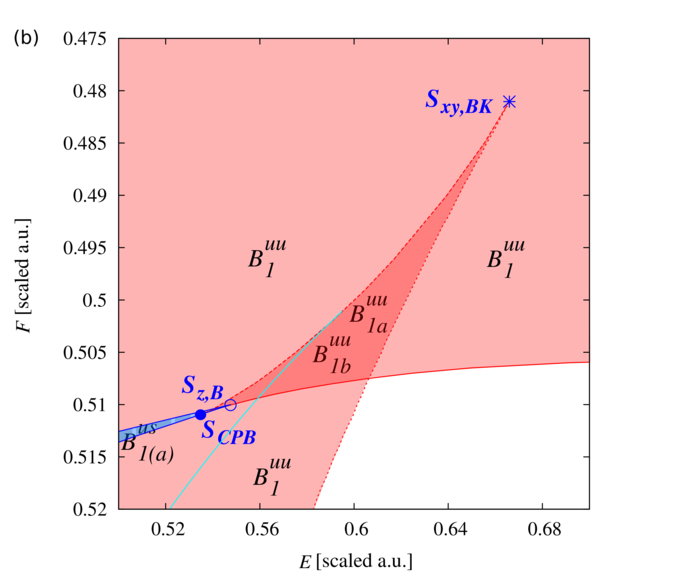}
\caption{(Color online) Stability of $B_{1}$ in dependence on the scaled 
energy and the scaled field strength. (a) is an enlargement of 
the rectangular area marked by dashed lines in Fig.~\ref{fig:B1B2}a.
(b) is an enlargement of the equivalently marked area in (a). 
The different resolutions allow for a comparison with
Fig.~\ref{fig:B2g} and provide a better insight in the relevant region
around $F\approx0.5,\, E\approx0.6$. In the dark red area two different
versions of $B_{1}$ exist\textcolor{black}{, which are indicated by $B_{1a}$ and $B_{1b}$. 
The field strength $F$ and the energy $E$ are given in scaled atomic units (see Eq.~(\ref{eq:skal})).}
\label{fig:B1g}}
\end{figure}

In the following figures we indicate the stability of 
orbits by two upper indices added to the name of the orbit.
The first one refers to the stability (s) or instability (u)
perpendicular to the magnetic field. The second one indicates the stability 
parallel to the magnetic field. Regions of different stability 
are additionally dyed by colors\textcolor{black}{, shadings} and grayscales.
The stability behavior of the orbits
$B_{1}$ is shown in Figure~\ref{fig:B1B2}a. 
After coming into existence they are at first stable parallel
to the magnetic field, since the Stark saddle point is a local potential
minimum in this direction. Towards higher energies they become completely
unstable and may vanish in bifurcations. We note that above a specific
field strength of $F_{xy,BK}=0.481$ the right point of intersection
with the $x$-axis of the orbits $B_{1}$ coincides with the nucleus.
This coincidence occurs along the \textcolor{black}{dash dotted black} line in Fig.~\ref{fig:B1B2}a, which 
passes through the lower area of instability.
Along this line the orbits $B_1$ are closed periodic orbits starting 
at and returning to the nucleus \citep{MainDu87, MainDe88c, MainBog88b}.
Afterwards a change of the shape takes place before the orbit vanishes
in a pitchfork bifurcation with period-doubling along the right dashed
line in Fig.~\ref{fig:B1B2}a. 

In crossed fields two orbits exist, which are localized
in the $z=0$-plane and which 
change over to the Kepler orbits \citep{PRA75_4,Diss} in the case of vanishing fields.
We will refer to them as $K_1$ and $K_2$.
The second orbit, $K_{2}$, distinguishing itself from the first one by
its sense of rotation around the nucleus, is not involved in any of
the bifurcations considered and therefore not shown in Fig.~\ref{fig:orbits}b.
The lastly mentioned dashed red line marks also a stability change
perpendicular to the magnetic field of $K_1$, which is related to 
a bifurcation between $B_{1}$ and $K_{1}$. 
Beside the bifurcation with $K_{1}$ two other 
bifurcations can be observed for $B_1$: At the stability
change in $z$-direction pitchfork bifurcations with period-doubling
occur, in which three-dimensional orbits are created. Along the
remaining \textcolor{black}{solid} \textcolor{black}{black} line at high energies separating the white area from the
area of instability \textcolor{black}{in Fig.~\ref{fig:B1B2}a} a tangent bifurcation with $B_{2}$ occurs.
This orbit has a similar shape as $B_{1}$, but exits only at
high energies and field strengths (Fig.~\ref{fig:B1B2}b). If one
takes a closer look at the stability areas according to the $z$-direction
one can find that they are ending in an \textcolor{black}{apex} not only for $B_{1}$ but
also for $B_{2}$. Both \textcolor{black}{apexes} meet at the point $\boldsymbol{S}_{z,B}=\left(F_{z,B},\, E_{z,B}\right)=\left(0.510,\,0.548\right)$
which is located on the bifurcation line between $B_{1}$ and $B_{2}$.
\textcolor{black}{Therefore we conclude that stability borders limiting
the area of stability in $z$-direction for $B_{1}$ continue as the according stability borders of $B_{2}$.}

\begin{figure}[t]
\includegraphics[width=0.49\columnwidth]{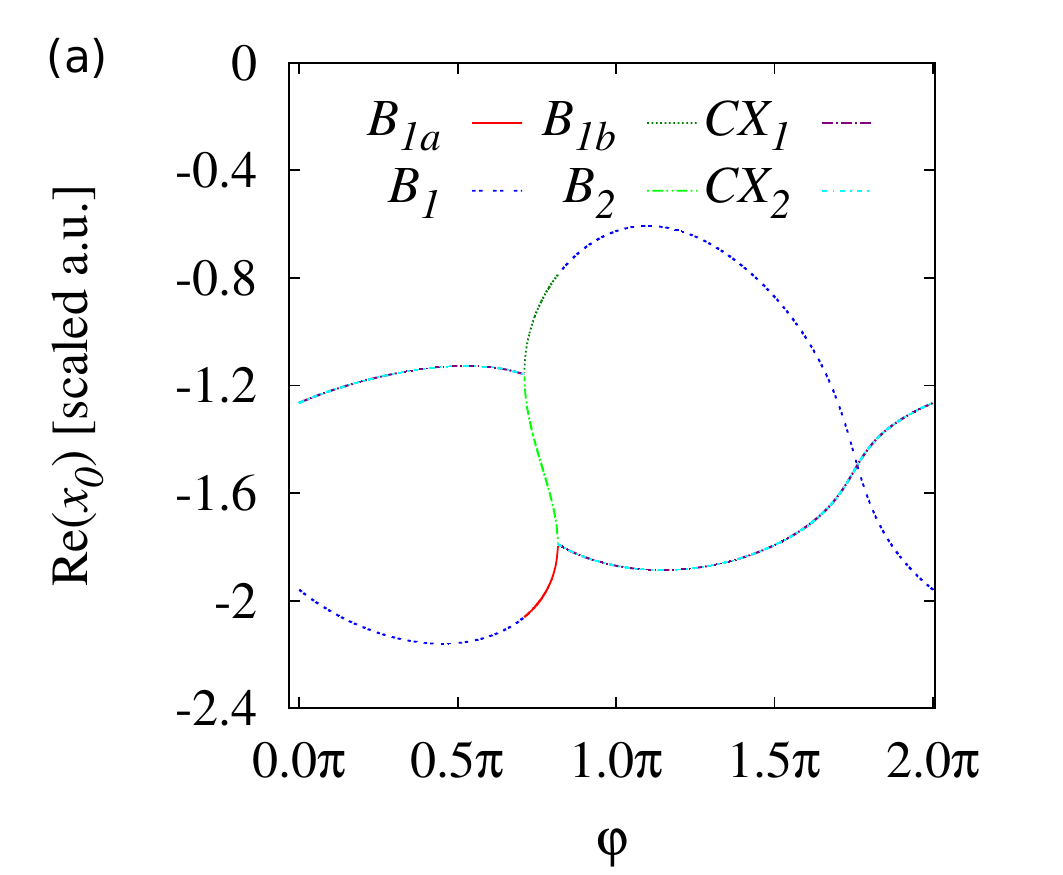}
\includegraphics[width=0.49\columnwidth]{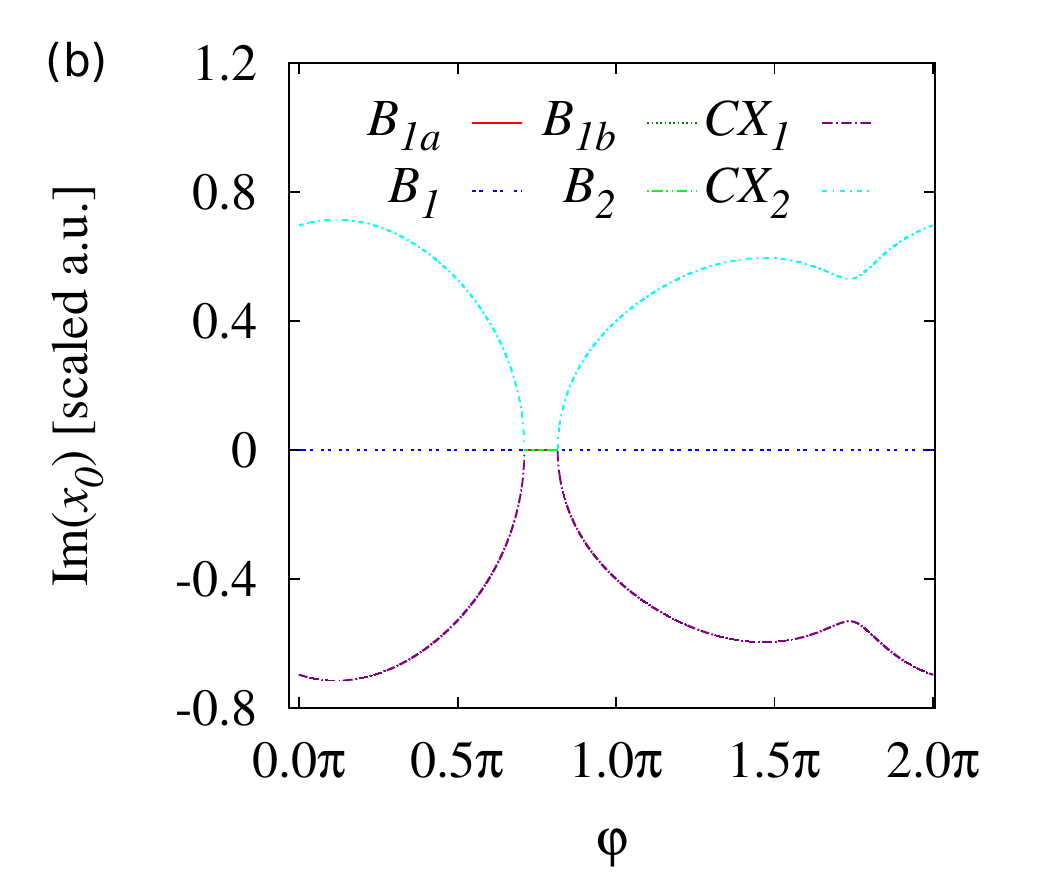}
\caption{(Color online) Real (a) and imaginary part (b) of one of the intersection points
of the orbits with the $x$-axis when encircling the cusp point $\boldsymbol{S}_{\mathrm{CP}B}$.
The angle $\varphi$ parameterizes the circle around $\boldsymbol{S}_{\mathrm{CP}B}$.
$CX_{1}$ and $CX_{2}$ are the two complex orbits appearing. Just
as in Fig.~\ref{fig:cusp}c and~\ref{fig:cusp}d it is not possible to determine the permutation behavior 
due to the coincidence of two orbits in a tangent bifurcation, respectively. 
\textcolor{black}{The angle $\varphi$ and the values of $x_0$ are given in dimensionless and scaled atomic units, respectively (see Eq.~(\ref{eq:skal})).}\label{fig:x0phi}}
\end{figure}

\begin{figure}[t]
\includegraphics[width=0.9\columnwidth]{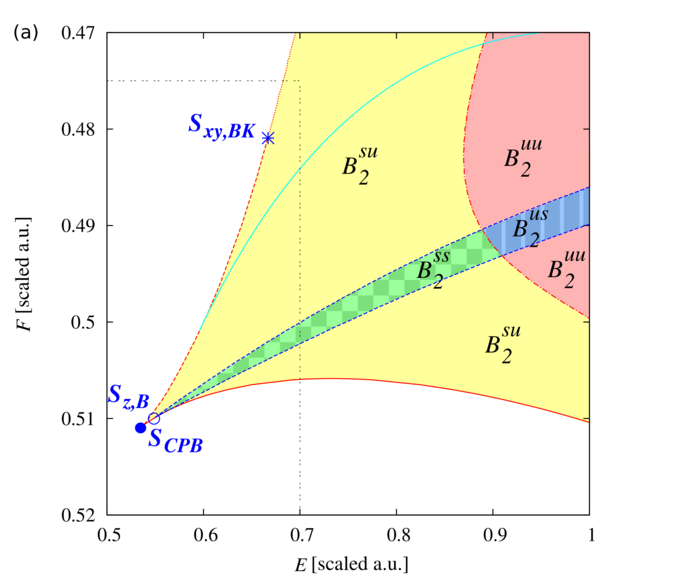}
\includegraphics[width=0.9\columnwidth]{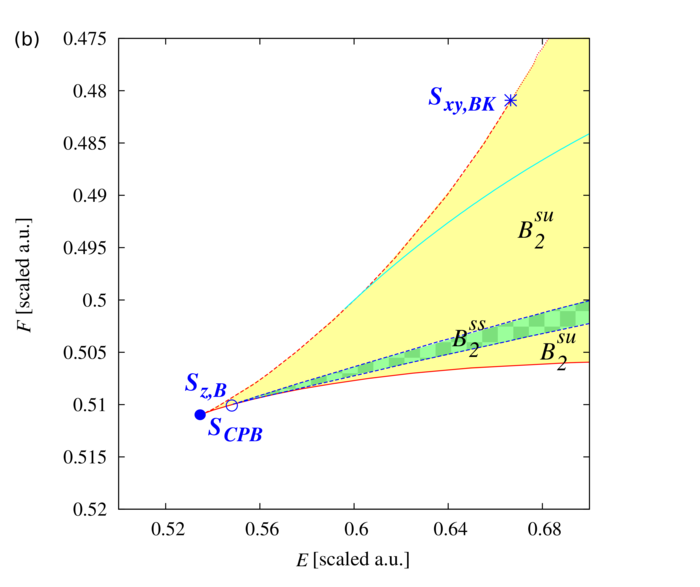}
\caption{(Color online) (a) Stability of $B_{2}$ in dependence on the scaled energy and the scaled
field strength. The observed stability island is triangular in
shape \textcolor{black}{indicated by $B_{2}^{ss}$} and ends in $\boldsymbol{S}_{z,B}$.
\textcolor{black}{(b) is an enlargement of the equivalently marked area in (a). 
The field strength $F$ and the energy $E$ are given in scaled atomic units (see Eq.~(\ref{eq:skal})).}
\label{fig:B2g}}
\end{figure}

Taking a closer look at this region of parameter space in Fig.~\ref{fig:B1g}b a cusp bifurcation
between orbits $B_{1}$ and $B_{2}$ is observed. As described in
Sec.~\ref{sub:Cusp-bifurcation-and} two tangent bifurcation lines
coincide in the cusp point --~here $\boldsymbol{S}_{\mathrm{CP}B}=\left(0.511,\,0.535\right)$~-- 
without continuation. Along both lines bifurcations
between $B_{1}$ and $B_{2}$ occur. But while $B_{2}$ exists only
in the area between both lines, $B_{1}$ exists in the complete external
region and in the area between both lines twice. 
\textcolor{black}{We will refer to those two different versions of $B_1$ as 
$B_{1a}$ and $B_{1b}$. The three orbits $B_{1a}$, $B_{1b}$ and $B_2$ correspond 
to the three fixed points of the cusp bifurcation in Sec.~\ref{sub:Cusp-bifurcation-and}.
Starting from the lower continuous line at the right-hand side of 
$\boldsymbol{S}_{\mathrm{CP}B}$ we follow the orbit $B_{1a}$ or $B_1$ anticlockwise around
the cusp point and again in the darker marked area until it vanishes as $B_{1b}$
in a tangent bifurcation along the dashed line between $\boldsymbol{S}_{\mathrm{CP}B}$
and $\boldsymbol{S}_{xy,BK}$. Therefore $B_{1a}$ and $B_{1b}$ can be converted
into each other by encircling $\boldsymbol{S}_{\mathrm{CP}B}$.} This behavior
is the same as for the fixed points of Eq.~(\ref{eq:cuspnormal}).
By allowing the coordinates and the time to become complex, the analytically continued orbits
can be followed beyond the tangent bifurcations. A plot of one of the
intersection points of the orbits with the $x$-axis vs. the angle
$\varphi$, which parameterizes the circle around $\boldsymbol{S}_{\mathrm{CP}B}$,
is shown in Fig.~\ref{fig:x0phi}.

The left tangent bifurcation line of $B_{1}$ and $B_{2}$ in Fig.
\ref{fig:B1g}b leading from $\boldsymbol{S}_{\mathrm{CP}B}$ to $\boldsymbol{S}_{xy,BK}$
coincides in $\boldsymbol{S}_{xy,BK}$ with the dashed pitchfork bifurcation
line of $B_{1}$ and $K_{1}$ \textcolor{black}{also ending in $\boldsymbol{S}_{xy,BK}$}. It continues itself towards lower field
strengths as a pitchfork bifurcation line between $B_{2}$ and $K_{1}$.
This line marks the upper boundary of the region in which $B_{2}$ exists.
Examinating the stability behavior of $B_{2}$ one can notice another
interesting phenomenon: The orbit becomes completely stable in a relatively
large area of parameter space (green area marked $B_2^{ss}$ in Fig.~\ref{fig:B2g}). This is used
as an opportunity to carry out a semiclassical quantizaton of $B_2$
and to search for signatures in exact quantum spectra in Sec.
\ref{sec:5}.

\begin{figure}[t]
\includegraphics[width=0.9\columnwidth]{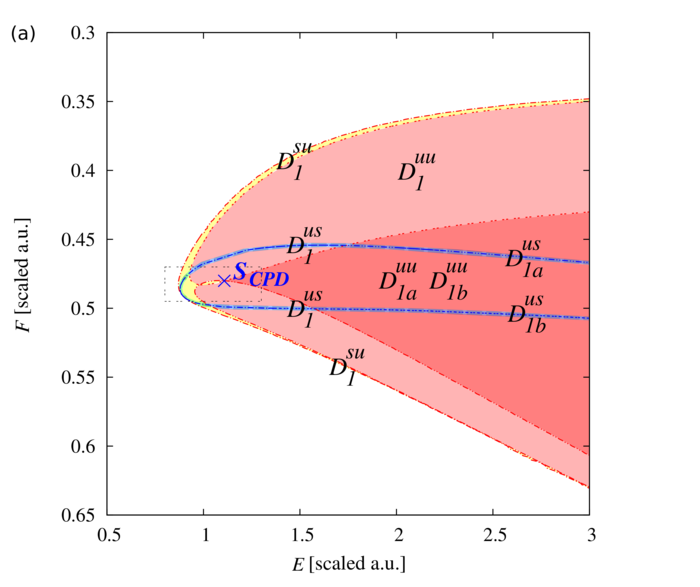}
\includegraphics[width=0.9\columnwidth]{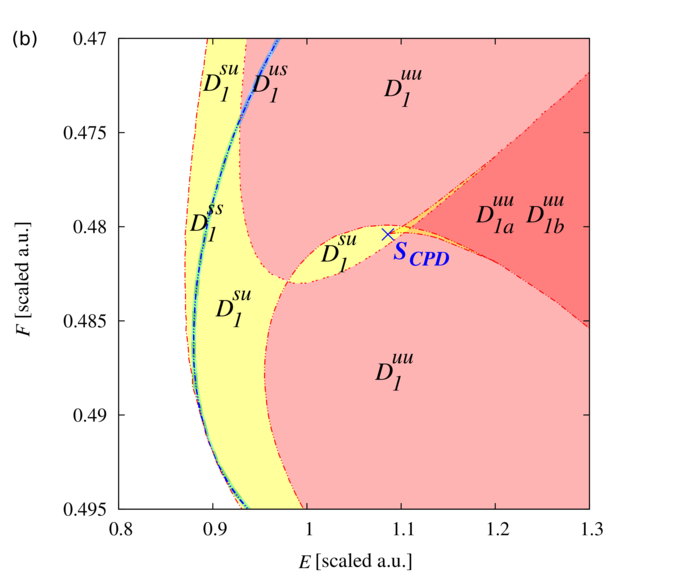}
\caption{(Color online) Stability of $D_{1}$ in dependence on the scaled energy and the scaled
field strength. A very \textcolor{black}{thin region marked $D_1^{ss}$} of complete stability can
be observed. A cusp bifurcation occurs around the cusp point
$\boldsymbol{S}_{\mathrm{CP}D}$.
\textcolor{black}{(b) is an enlargement of the equivalently marked area in (a). 
The field strength $F$ and the energy $E$ are given in scaled atomic units (see Eq.~(\ref{eq:skal})).}
\label{fig:D2}}
\end{figure}

Finally one last orbit localized in the $z=0$-plane is worth mentioning. 
Along the \textcolor{black}{dash dotted} \textcolor{black}{black} line in Fig.~\ref{fig:B1B2}b describing
a stability change of $B_{2}$ perpendicular to the magnetic field
a pitchfork bifurcation with period-doubling occurs. The period-doubled
orbit coming into being is named $D_{1}$ (compare Fig.~\ref{fig:orbits}). For $D_{1}$ both of the
described phenomena can be found again! In Fig.~\ref{fig:D2}b a very
small region of complete stability localized at even higher energies
than the stability island of $B_{2}$ is displayed. The cusp bifurcation
occurs around the cusp point $\boldsymbol{S}_{\mathrm{CP}D}$, in which
again two tangent bifurcation lines coincide without continuation.
The darker marked areas denote the existence of two different versions
of $D_{1}$: $D_{1a}$ and $D_{1b}$. We conclude that the reappearance
of those phenomena indicates the possibility to find them for other
orbits of even more complicated shape on-and-off-again.

\section{Semiclassical quantization and exact quantum-mechanical calculations\label{sec:4}}

According to periodic orbit theory every classical periodic orbit
causes a modulation in the density of states \citep{SemPhys,SemPhys1,MainBal70,MainBal71,MainBal72,MainBal74,MainGut67,MainGut69,MainGut70,MainGut82}.
Oscillating modulations can also be found in photoabsorption spectra
or action spectra, respectively, and are according to Gutzwiller \citep{SemPhys1}
and Miller \citep{Miller} located at
\begin{equation}
S-\sum_{i=1}^{2}\left(m_{i}+\frac{1}{2}\right)\varphi_{i}=2\pi\left(n+\frac{\lambda}{4}\right),
\label{eq:semclass}
\end{equation}
with the action $S$, the stability angles $\varphi_{i}=\mathrm{arg}\left(d_{i}\right)$,
determined by the eigenvalues $d_{i}$ with $\left|d_i\right|\geq1$ of the monodromy matrix, indicating
the stability parallel $\left(\parallel\right)$ and perpendicular
$\left(\perp\right)$ to the magnetic field, the Maslov index $\lambda$
and the quantum numbers $n,\, m_{1},\, m_{2}$. Those quantum numbers
count the number of quanta along $\left(n\right)$ and perpendicular
$\left(m_{i}\right)$ to the periodic orbit. Stable periodic orbits
cause modulations of the density of states, which can be described
by $\delta$-function peaks, while unstable orbits cause broadened
peaks. \textcolor{black}{Those broadened peaks have the shape of 
Lorentzians in action spectra, i.e., when they are plotted against 
the action $S$. In the following we will use the term semiclassical 
half-width for the half-widths of those peaks according to energy. For each value 
of $n$ and each constant values of the field strengths the widths 
according to energy can be calculated as the difference 
between the two points, where 
\begin{equation}
S\pm\sum_{i=1}^{2}\mathrm{ln}\left|d_{i}\right|-\sum_{i=1}^{2}\left(m_{i}+\frac{1}{2}\right)\varphi_{i}=2\pi\left(n+\frac{\lambda}{4}\right)
\label{eq:semclass2}
\end{equation}
is fulfilled \citep{SemPhys1}.} Since
we did not calculate the Maslov index of $B_{2}$, we assume in the
following $\lambda=1$, which will supply the expected agreement between
semiclassical and quantum-mechanical results.

\begin{figure}[t]
\includegraphics[width=0.95\columnwidth]{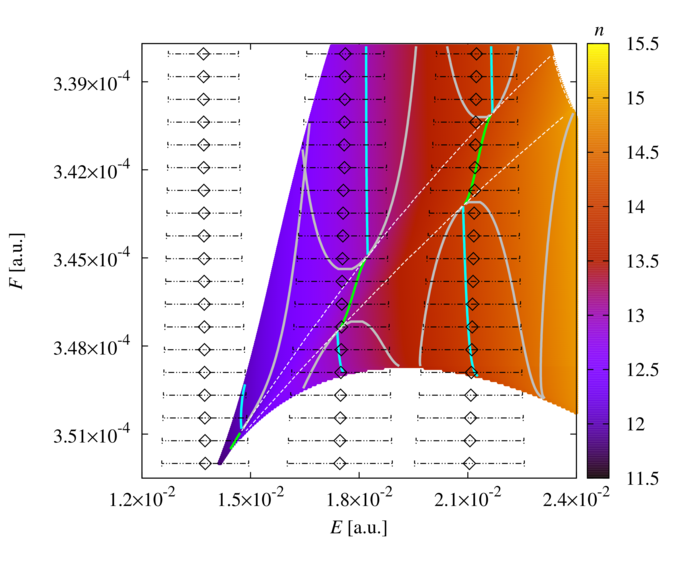}
\includegraphics[width=0.95\columnwidth]{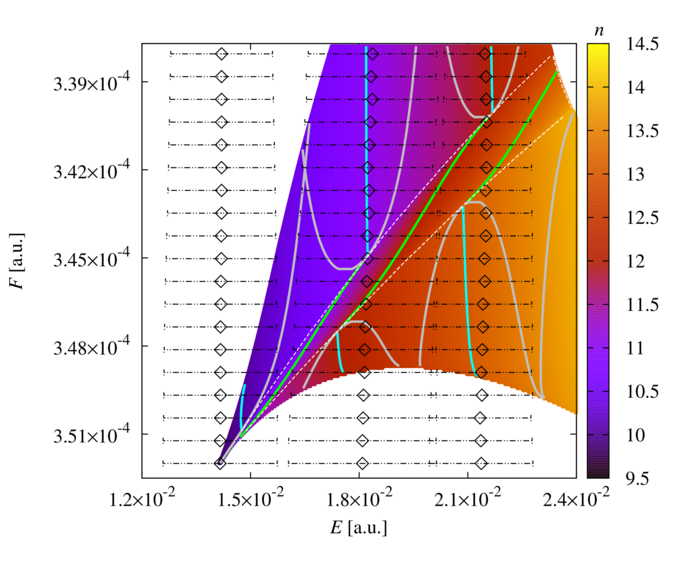}
\caption{
(Color online) Comparison between quantum-mechanical and semiclassical results at
$B=4.26\times10^{-3}\left(\hat{=}1000\,\mathrm{T}\right)$. \textcolor{black}{For further information see text of Sec.~\ref{sec:5}.} Best
agreements are obtained for $\lambda=1$ and $m_{\perp}=0,\, m_{\parallel}=0$
in case of the first resonance series (a) and $m_{\perp}=0,\, m_{\parallel}=2$
in case of the second resonance series (b).
\textcolor{black}{The parameter range shown is almost the same as in Fig.~\ref{fig:B2g}a. All values are given in atomic units.}
\label{fig:Sem1000}
}
\end{figure}

\begin{figure}[t]
\includegraphics[width=0.95\columnwidth]{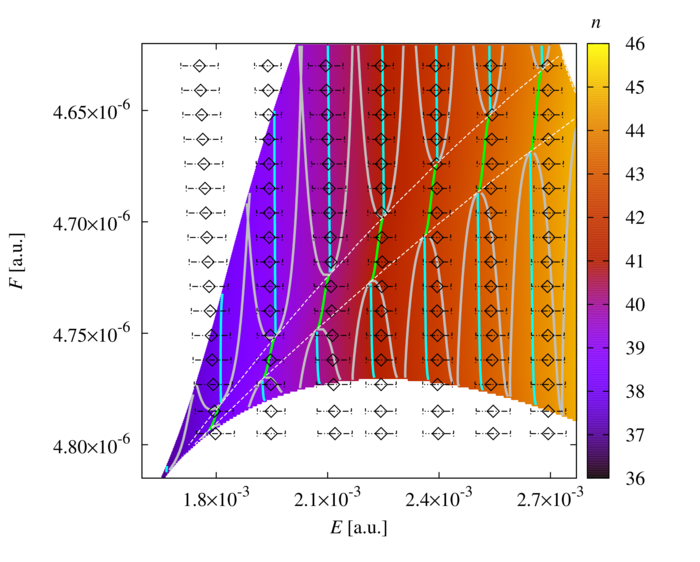}
\caption{
(Color online) Comparison between semiclassical and quantum-mechanical results for
the first resonance series at $B=1.70\times10^{-4}\left(\hat{=}40\,\mathrm{T}\right)$. The resonance widths
are evidently smaller but still do not vanish within the stability
island, which is marked by dashed lines. \textcolor{black}{For further information see text of Sec.~\ref{sec:5}.} 
\textcolor{black}{All values are given in atomic units.}
\label{fig:Sem40}
}
\end{figure}

The quantum-mechanical resonance spectra and wave functions are determined
as described in \citep{PRA81}: The Schr\"odinger equation with the \textcolor{black}{unscaled}
Hamiltonian (\ref{eq:H}) is rewritten in dilated semiparabolic coordinates
\citep{PRA81_13,PRA79}
\begin{equation}
\mu=\frac{1}{b}\sqrt{r+z},\;\nu=\frac{1}{b}\sqrt{r-z},\;\mathrm{and}\;\varphi=\arctan\frac{y}{x}.
\end{equation}
\textcolor{black}{Note that we cannot further use scaled variables in quantum mechanical calculations since the scaling is restricted to the classical dynamics \citep{MainHar83,Diss}.}
The parameter $b=\left|b\right|e^{i\vartheta/2}$ introduces a complex
delatation of the coordinates $\boldsymbol{r}$, which is necessary to
determine resonances using the complex coordinate rotation method
\citep{PRA81_14,PRA81_15,PRA81_16}. The Schr\"odinger equation then
reads
\[
\left[\triangle_{\mu}+\triangle_{\nu}-\left(\mu^{2}+\nu^{2}\right)+4b^{2}+b^{4}B\left(\mu^{2}+\nu^{2}\right)i\frac{\partial}{\partial\varphi}\right.
\]
\[
\left.-\frac{1}{4}b^{8}B^{2}\mu^{2}\nu^{2}\left(\mu^{2}+\nu^{2}\right)-2b^{6}F\mu\nu\left(\mu^{2}+\nu^{2}\right)\cos\varphi\right]\psi
\]
\begin{equation}
=\left[\Lambda\left(\mu^{2}+\nu^{2}\right)\right]\psi,\label{eq:11}
\end{equation}
with the Laplacians
\begin{equation}
\triangle_{\rho}=\frac{1}{\rho}\frac{\partial}{\partial\rho}\rho\frac{\partial}{\partial\rho}+\frac{1}{\rho^{2}}\frac{\partial^{2}}{\partial\varphi^{2}},\;\rho\in\left\{ \mu,\,\nu\right\} ,
\end{equation}
and generalized complex eigenvalues $\Lambda=-\left(1+2b^{4}E\right)$,
related to the complex energies $E$ of resonances, the real parts
of which representing their energies and the imaginary parts their
widths $\Gamma=-2\mathrm{Im}\left(E\right)$. To calculate resonances
a matrix representation of the Schr\"odinger equation~(\ref{eq:11})
is diagonalized. We use the adequate complete basis
\begin{equation}
\left|n_{\mu},\, n_{\nu},\, m\right\rangle =\left|n_{\mu},\, m\right\rangle \otimes\left|n_{\nu},\, m\right\rangle ,
\end{equation}
with the eigenstates $\left|n_{\rho},\, m\right\rangle $ of the two-dimensional
harmonic oscillator. The position space representation reads
\[
\psi_{n_{\mu}n_{\nu}m}\left(\mu,\,\nu,\,\varphi\right)=\left\langle \mu,\,\nu,\,\varphi\left|n_{\mu},\, n_{\nu},\, m\right.\right\rangle
\]
\[
 =\sqrt{\frac{\left[\left(n_{\mu}-\left|m\right|\right)/2\right]!\left[\left(n_{\nu}-\left|m\right|\right)/2\right]!}{\left[\left(n_{\mu}+\left|m\right|\right)/2\right]!\left[\left(n_{\nu}+\left|m\right|\right)/2\right]!}}
\]
\begin{equation}
\times\sqrt{\frac{2}{\pi}}f_{n_{\mu}m}\left(\mu\right)f_{n_{\nu}m}\left(\nu\right)e^{im\varphi},
\end{equation}
with
\begin{equation}
f_{nm}\left(\rho\right)=e^{-\rho^{2}/2}\rho^{\left|m\right|}\mathcal{L}_{\left(n-\left|m\right|\right)/2}^{\left|m\right|}\left(\rho^{2}\right),
\end{equation}
and the associated Laguerre polynomials $\mathcal{L}_{k}^{\alpha}\left(x\right)$.
In our numerical calculations the maximum number of states used is
limited by the condition $n_{\mu}+n_{\nu}\leq60$ due to the required
computer memory. Since $\mu$ and $\nu$ are complex coordinates,
all non-intrinsically complex parts must remain unconjugated in case
of a complex conjugation \citep{PRA81_16,PRA81_17}. This means that
$\psi_{n_{\mu}n_{\nu}m}^{*}$ is equal to $\psi_{n_{\mu}n_{\nu}m}$
except for the term $e^{im\varphi}$ which is replaced with $e^{-im\varphi}$.
The eigenstates
\begin{equation}
\Psi_{i}\left(\mu,\,\nu,\,\varphi\right)=\sum_{n_{\mu},\, n_{\nu},\, m}c_{i\, n_{\mu},\, n_{\nu},\, m}\psi_{n_{\mu}n_{\nu}m}\left(\mu,\,\nu,\,\varphi\right),
\end{equation}
\begin{equation}
\Psi_{i}^{*}\left(\mu,\,\nu,\,\varphi\right)=\sum_{n_{\mu},\, n_{\nu},\, m}c_{i\, n_{\mu},\, n_{\nu},\, m}\psi_{n_{\mu}n_{\nu}m}^{*}\left(\mu,\,\nu,\,\varphi\right),
\end{equation}
obtained as a result of the matrix diagonalization, are normalized
according to
\[
\int\,\mathrm{d}^{3}\boldsymbol{r}\,\Psi_{i}^{*}\left(\mu,\,\nu,\,\varphi\right)\Psi_{j}\left(\mu,\,\nu,\,\varphi\right)
\]
\begin{equation}
=b^{6}\int_{0}^{\infty}\,\mathrm{d}\mu\,\int_{0}^{\infty}\,\mathrm{d}\nu\,\int_{0}^{2\pi}\,\mathrm{d}\varphi\,\mu\nu\left(\mu^{2}+\nu^{2}\right)\Psi_{i}^{*}\Psi_{j}=\delta_{ij}.
\end{equation}

\begin{figure}[t]
\includegraphics[width=0.9\columnwidth]{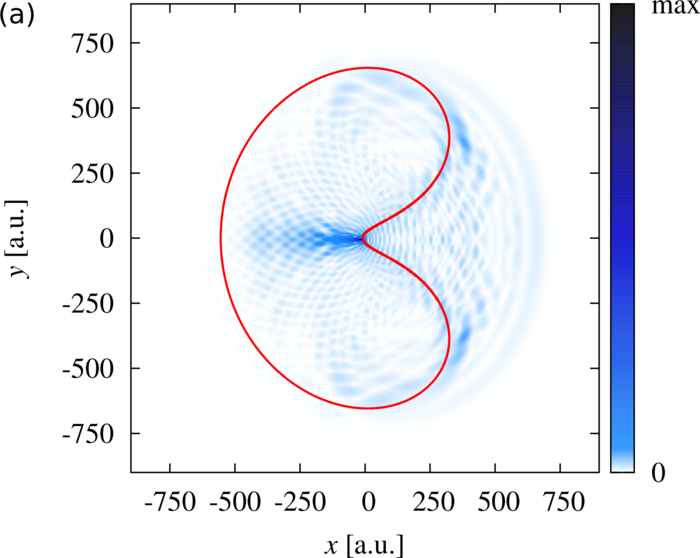}

\includegraphics[width=0.9\columnwidth]{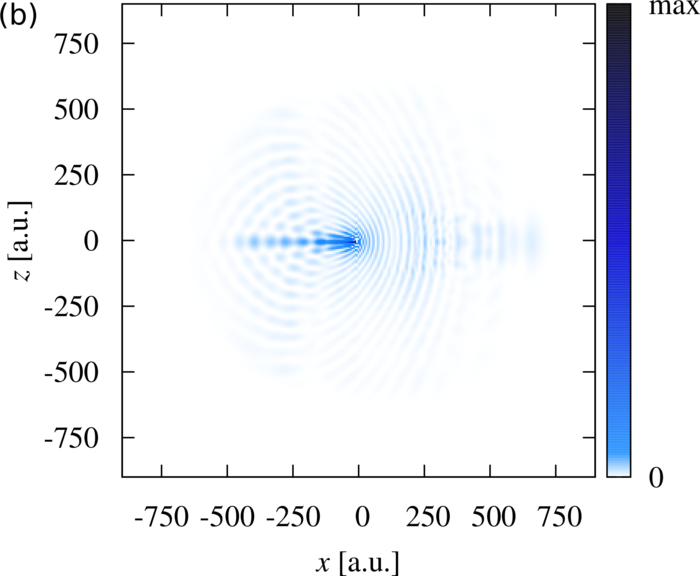}

\caption{
(Color online) Probability density $\varrho\left(\boldsymbol{r}\right)$ of a resonance
of the first series at $E=1.95\times10^{-3}-4.14\times10^{-5}i$,
$F=4.76\times10^{-6}$ and $B=1.70\times10^{-4}$. One can observe an agreement with the course
of the classical orbit (continuous line) in (a) and the restriction
to the $z=0$-plane in (b). 
\textcolor{black}{All values are given in atomic units.}
\label{fig:WF1}
}
\end{figure}

\begin{figure}[t]
\includegraphics[width=0.9\columnwidth]{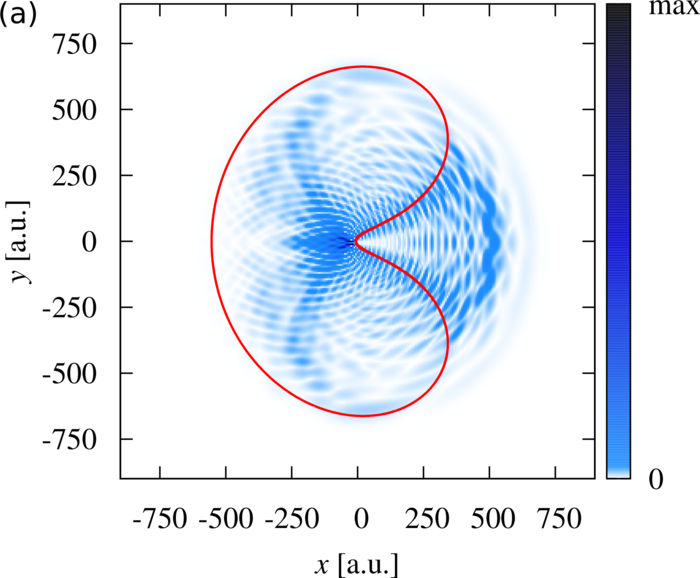}

\includegraphics[width=0.9\columnwidth]{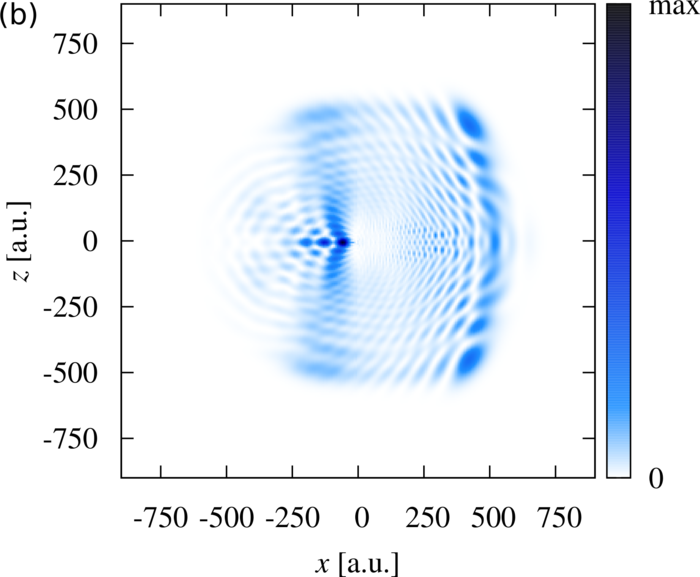}

\caption{
(Color online) Probability density $\varrho\left(\boldsymbol{r}\right)$ of a resonance
of the second series at $E=2.14\times10^{-3}-5.87\times10^{-5}i$,
$F=4.73\times10^{-6}$ and $B=1.70\times10^{-4}$. The differences to the classical orbit (continuous
line) are significantly larger. 
\textcolor{black}{All values are given in atomic units.}
\label{fig:WF2}
}
\end{figure}

Due to the restriction of the complex conjugation to the intrinsically
complex parts the expression $\Psi_{j}^{*}\Psi_{j}$ is not a real
quantity. The probability density is obtained as $\varrho=\left|\Psi_{j}^{*}\Psi_{j}\right|$
\citep{NHQM} instead. Finally Husimi distributions can be calculated,
allowing a comparison between the classical torus structure existing
around the periodic orbits and quantum-Poincar\'{e} sections \citep{DissMueller_Leb89}.
To prevent a divergence of the momenta we use regularized coordinates
for the Husimi distributions, too. We will show that the probability
density of the observed resonances is mainly limited to the $z=0$-plane.
For that reason the calculations can be restricted to the $x$- and
$y$- or $U_{1}$- and $U_{2}$-coordinate, respectively (cf.~Eq.~(\ref{eq:reg_r_U}) and~(\ref{eq:U_r})). 
By setting $z=0$ one obtains
\begin{equation}
\left(\begin{array}{c}
x\\
y
\end{array}\right)=\left(\begin{array}{c}
U_{1}^{2}-U_{2}^{2}\\
2U_{1}U_{2}
\end{array}\right),
\end{equation}
which is a bijection assuring $\alpha=0$ as long as $U_{2}\geq0$
holds. The Husimi distribution \citep{PRL55_4,PRL55_5,PRL55_6,TsaiDuke_Bal98}
then reads
\[
P_{H}\left(\boldsymbol{U},\,\boldsymbol{P}\right)=\left(\int\,\mathrm{d}^{2}\boldsymbol{\Xi}\,4\left|\boldsymbol{\Xi}\right|^{2}\Psi\left(\boldsymbol{\Xi}\right)G\left(\boldsymbol{\Xi},\,\boldsymbol{U},\,\boldsymbol{P}\right)\right)
\]
\begin{equation}
\times\left(\int\,\mathrm{d^{2}}\boldsymbol{\Xi}\,4\left|\boldsymbol{\Xi}\right|^{2}\Psi^{*}\left(\boldsymbol{\Xi}\right)G^{*}\left(\boldsymbol{\Xi},\,\boldsymbol{U},\,\boldsymbol{P}\right)\right),
\end{equation}
with a Gaussian of minimum uncertainity $\left(\Delta U\Delta P={1/2}\right)$
\begin{equation}
G\left(\boldsymbol{\Xi},\,\boldsymbol{U},\,\boldsymbol{P}\right)=\frac{1}{\left(\sigma^{2}\pi\right)^{{1/2}}}e^{-\frac{1}{2\sigma^{2}}\left(\boldsymbol{\Xi}-\boldsymbol{U}\right)^{2}-i\left(\boldsymbol{P}+\boldsymbol{A}\left(\boldsymbol{U}\right)\right)\boldsymbol{\Xi}},
\end{equation}
the vector potential 
\begin{equation}
\boldsymbol{A}=\boldsymbol{P}+\frac{1}{2}B\left(U_{1}^{2}+U_{2}^{2}\right)\left(\begin{array}{c}
-U_{2}\\
U_{1}
\end{array}\right),
\end{equation}
and an appropriate adaptable squeezing parameter $\sigma=\sqrt{\Delta U/\Delta P}$.
To determine quantum-Poincar\'{e} sections $U_{1}$ is set to the
constant value $U_{1}=0$. The conjugate momentum $P_{1}$ is then
calculated according to Eq.~(\ref{eq:reg_H}).

\section{Results of quantum-mechanical calculations and discussion \label{sec:5}}

\begin{figure}[t]
\includegraphics[width=0.9\columnwidth]{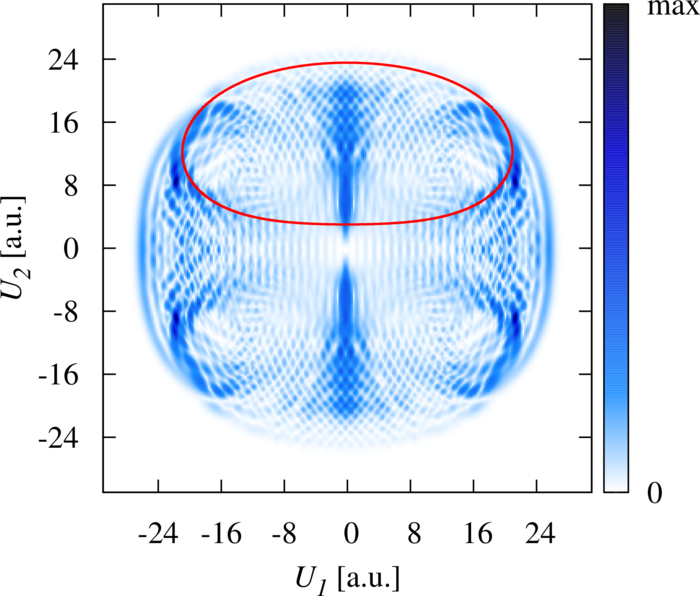}

\caption{
(Color online) Probability density $\left|4U^{2}\Psi^{*}\left(\boldsymbol{U}\right)\Psi\left(\boldsymbol{U}\right)\right|$
of the resonance of Fig.~\ref{fig:WF1} in regularized coordinates.
Without the limitation of $U_{2}\geq0$ the probability density shows
a second symmetry relative to the line $U_{2}=0$. The lower part
$\left(U_{2}\leq0\right)$ corresponds to $\alpha=2\pi$ and is not
included in the calculations of the quantum-Poincar\'{e}
sections. The classical orbit intersects the $U_{1}=0$-line twice
but with a different sign of the velocity $V_{1}$ in $U_{1}$-direction.
\textcolor{black}{All values are given in atomic units.}
\label{fig:WFreg}
}
\end{figure}
\begin{figure}[t]
\includegraphics[width=0.95\columnwidth]{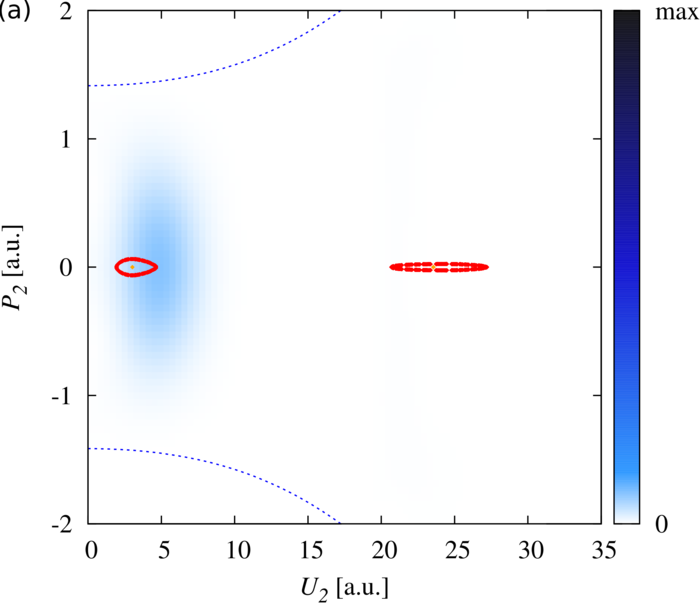}

\includegraphics[width=0.95\columnwidth]{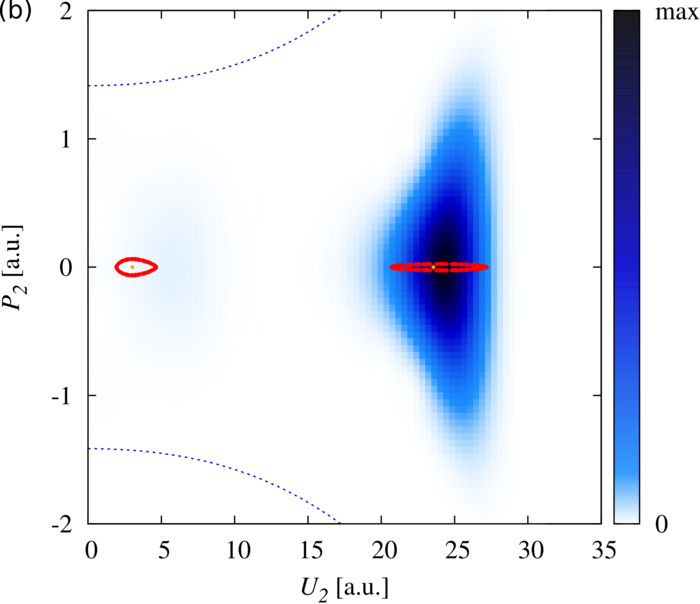}

\caption{(Color online) Comparison between quantum-Poincar\'{e}
sections (shown by colorbar (grayscales)) and classical torus structure. The dashed
lines display the borders of the classically permitted region. Since
the Hamiltonian is a quadratic function of the momentum, there are
two possibilities to choose $P_{1}$ when calculating the section.
Those both solutions differ from each other in the sign of the velocity
$V_{1}$ in $U_{1}$-direction. Therefore agreements with only one
part of the torus structure can be achieved at any time. The colorbar
is chosen equal to compare both results. We set $\sigma$ to the constant
value $\sigma=2$.
\textcolor{black}{All values are given in atomic units.}
\label{fig:Husimi}}
\end{figure}

Since the classical calculations have been carried out for scaled energy
values and scaled field strengths one of the three parameters $E$,
$F$ and $B$ can now be chosen arbitrarily in quantum-mechanical
calculations. In the following we will set $B$ to fixed values while
the other parameters are calculated according to Eq.~(\ref{eq:skal})
and the location of the stability island. Figure~\ref{fig:Sem1000}
shows results at $B=\unit[1000]{T}$. 
\textcolor{black}{The colorbar (grayscales) of this figure shows the quantum numbers $n$
obtained by semiclassical quantization of $B_{2}$ according to Eq.~(\ref{eq:semclass}).
The stability island is marked by dashed lines. Green (solid) lines within the stability island and 
blue (solid) lines outside of it display positions where integer values of $n$ are obtained. 
White (solid) lines of parabolic shape outside the stability island display the semiclassical 
half-widths calculated by Eq.~(\ref{eq:semclass2}). 
For each constant value of $F$ the semiclassical half-width (according to energy) of a semiclassical state with quantum number $n$ can be read out as the sum of the distances between the blue (solid) line for this value of $n$ and the neighboring white (solid) lines.
The real parts of the quantum-mechanical resonances (calculated only at discrete field strengths)
are displayed by diamonds, the imaginary parts by error bars.}

Two series of resonances
were found, which can be traced down to lower energies. It is supposed
that this retracing will end in the resonances described in \citep{PRA81}.
\textcolor{black}{The contradictory fact that the resonances can also be found 
outside the region where the classical orbits $B_2$ exist is an indication 
that a simple semiclassical quantization including solely $B_2$ may be 
insufficient (cf.\ also Sec.~\ref{sec:6}). Semiclassical quantization normally includes all orbits 
appearing at a specific point in parameter space \citep{SemPhys}. 
However, we assume that the resonances found within the 
area of $B_2$ are closely related to those orbits.}

The energetic distance between two resonances according to the real part of the energy is
almost the same within both series while the widths differ strongly
between the two series. Assuming that the resonances of smaller widths
(first series) represent a quantum-mechanical ground state and the
other ones (second series) an exited state, semiclassical calculations
are carried out to the quantum numbers $m_{\perp}=0,\, m_{\parallel}=0$
and $m_{\perp}=0,\, m_{\parallel}=2$, respectively. Since we consider
only states of even parity in our quantum-mechanical calculations,
the state with $m_{\perp}=0,\, m_{\parallel}=2$ must be the first
excited state appearing in the spectra. As concerns the positions
of the resonances a very good agreement between semiclassical and
quantum-mechanical results is achieved \textcolor{black}{outside the stability island. 
However, those quantum numbers lead to a switching of the semiclassical 
results from one resonance to another. Using different quantum numbers an 
even worse agreement outside the stability island is obtained. We therefore 
conclude that $m_{\perp}=0$ and $m_{\parallel}=2$ are correct and that the 
quantum mechanical resonances do not show the switching behavior since they are not 
able to resolve the semiclassical structure in this region, which we will explain in the following.
On the other hand this switching can be seen as a further hint that a simple semiclassical 
quantization of $B_2$ is insufficient and that all the other orbits have to be considered as well.}

\textcolor{black}{As can be seen from Fig.~\ref{fig:Sem1000}, a good 
agreement according to the widths of the resonances is 
achieved only far away from the stability island.} The expected decline
of resonance widths within the stability island, proving the existence
of quantum-mechanical bound states related to the classical stable
orbits $B_{2}$, is not observed and will only show up at lower magnetic
field strengths. When calculating Poincar\'{e} sections of the classical
torus structure existing around the periodic orbit one uncovers that
this structure is very small, too small to be resolved in quantum-mechnical
calculations at $B=\unit[1000]{T}$.

Due to the expansion of this structure with a decreasing value of
$B$ (note that $\boldsymbol{r}=\tilde{\boldsymbol{r}}B^{-2/3}$)
one expects a better resolution at lower field strengths. The increasing
quantum number of $n$ makes it necessary to increase the number of
basis functions used in order to ensure a convergence of the quantum-mechanical
calculations. Within the limitation of $n_{\mu}+n_{\nu}\leq60$ it
is possible to follow the resonances down to $B=\unit[40]{T}$. Figure~\ref{fig:Sem40} 
shows that even at this field strength the resonance
widths do not vanish within the stability island. The probability
densities (Fig.~\ref{fig:WF1} and~\ref{fig:WF2}) exhibit instead
a partially good agreement with the course of the classical orbit.
Differences can be explained by the fact that the calculated wave
function is only a snapshot while the resonance itself is a time evolving
and finally decaying state. 
\textcolor{black}{The higher quantum mechanical probability density 
along the negative $x$ axis possibly indicates the decay of the resonance 
since -- from a classical point of view -- all trajectories of the electron 
describing the ionization of the hydrogen atom have to pass the vicinity of 
the Stark saddle point \citep{PRA81}. This higher probability density then 
shows up again along the $U_2$ axis in Fig.~\ref{fig:WFreg} (and will finally 
be seen as a displacement between classical and quantum mechanical structure in Fig.~\ref{fig:Husimi}a.}
It can be observed that the larger values
of the probability density are taken on in the $z=0$-plane when regarding
the first resonance series. In case of the second resonance series
the extension in the $z$-direction is much larger. 

The calculated quantum-Poincar\'{e} sections are compared with the
classical ones in Fig.~\ref{fig:Husimi}. As expected, the torus structure
is, especially in direction of the momentum $P_{2}$, much smaller than
the quantum-mechanical structure.

Finally we want to estimate at which field strengths
a solution of the classical structure may be possible. Since the classical
orbit is localized in the $z=0$-plane we regard the extension of
the torus structure around the orbit in a $z$-$p_{z}$-Poincar\'{e}
section as its classical uncertainity \textcolor{black}{(note that we 
change back to non-regularized coordinates in this place).}
Assuming that the quantum-mechanically
determined uncertainity product $\Delta z\Delta p_{z}=1.59$ for one
resonance of the first series does not change to a great extent with
the value of $B$, since it is already sufficiently close to the critical
value of $\frac{1}{2}$, we obtain a coincidence not until $B\approx\unit[90]{mT}$!
This field strength is convenient for an experimental proof of the
orbits $B_{2}$ in photoabsorption spectra. As it has been described, one 
of the intersection points of the classical orbits with the $x$-axis
is very close to the nucleus within the relevant region around 
$F\approx0.5,\, E\approx0.6$. An exisiting overlap of the corresponding 
resonances and the lowest states (1s or 2p) of the hydrogen atom allows 
therefore an experimental excitation of the hydrogen atom into those 
resonance states. According to closed orbit theory 
\citep{MainDu87, MainDe88c, MainBog88b} and periodic orbit 
theory \citep{MainBal70, MainBal71, MainBal72, MainBal74, 
MainGut67, MainGut69, MainGut70, SemPhys1, MainGut82} the classical 
periodic orbits then become noticeable in photoabsorption spectra by 
oscillating terms of the form 
$A\sin(2\pi\tilde{S}B^{-1/3}+\varphi)$, in which the amplitudes depend 
on the stability of the orbits as well as the process of excitation.
For the purpose of an experimental proof the data
of an classical orbit located in the center of the stability island
is given in scaled atomic units: $\tilde{F}=0.4985$, $\tilde{E}=0.75$,
$\tilde{S}=14.7843$. The conversion to unscaled units at the chosen
magnetic field strength is obtained by Eq.~(\ref{eq:skal}) and $\tilde{S}=SB^{1/3}$.

\section{Summary and conclusion \label{sec:6}}

Following the quasi-Penning resonances up to high energies and field
strengths we could resolve their complete bifurcation behavior. We
found out that only $K_{1}$, $B_{2}$ and three-dimensional orbits
are involved in the bifurcations of $B_{1}$. It was shown that several
phenomena appear in the region of $F\approx0.5,\, E\approx0.6$ including
a cusp bifurcation between $B_{1}$ and $B_{2}$ as well as the appearance
of a stability island for $B_{2}$. The reappearance of those phenomena
when analysing the stability and bifurcation behavior of $D_{1}$
could then be interpreted as a possibility to find them on-and-off-again
for orbits of even more complicated shape. The puzzle of the $z$-stability
region of $B_{1}$ ending in an \textcolor{black}{apex} in parameter space could be resolved
by a closer examination of the stability of $B_{2}$. The results
show a coincidence of two stability \textcolor{black}{apexes} at $\boldsymbol{S}_{z,B}$,
which indicates the continuation of the stability borders limiting
the areas of stability in $z$-direction from one of those orbits
to the other one. 

Semiclassical quantizations of $B_{2}$ showed agreements with resonances
found in exact quantum spectra. Due to further agreements between
quantum-mechanical probability densities and classical orbits as well
as between quantum-Poincar\'{e} sections and classical torus structure
we have to conclude that signatures of the stable orbit $B_{2}$ in
exact quantum spectra have been found. Owing to limited computer memory
and power it is not yet possible to reach regions of several millitesla
in order to find out whether the widths of the detected resonances
will disappear within the stability island or not. Nevertheless we
think that an experimental proof of $B_{2}$ in photoabsorption spectra
at the estimated field strengths will eventually be possible.

Finally, since the widths of the observed resonances do not vanish
within the stability island of $B_{2}$ and since the stability island
is very close to the bifurcation lines with $B_{1a}$ and $B_{1b}$,
we think that the orbits $B_{2}$ cannot be treated as isolated ones
in a semiclassical quantization. It would be therefore preferable
to perform a uniform semiclassical quantization \citep{Ghost1,Ghost2,Ghost3}
of the observed cusp bifurcation.

\acknowledgments
We thank Eugen Fl\"othmann for stimulating discussions.



\end{document}